\documentstyle [prl,aps]{revtex}

\begin{document}

%\twocolumn[\hsize\textwidth\columnwidth\hsize\csname @twocolumnfalse\endcsname
\title{Friedmann Equation and
Stability of Inflationary Higher Derivative Gravity}

\author{ W.F. Kao\thanks{email:wfgore@cc.nctu.edu.tw},
Ue-Li Pen\thanks{email:pen@cita.utoronto.ca}, and Pengjie
Zhang\thanks{email:zhangpj@cita.utoronto.ca}}
\address{$^*$ Institute of Physics, Chiao Tung
University, Hsinchu, Taiwan\\
${}^\dagger$
Canadian Institute for Theoretical Astrophysics, University of
Toronto, Toronto, Canada\\
${}^\ddagger$ Astronomy Department, University of Toronto, Toronto, Canada }
%\date{Jan. 2001, Revised}
\maketitle

% Abstract
\begin{abstract}
Stability analysis on the De Sitter universe in pure gravity theory is known to be useful in many aspects.
We first show how to complete the proof of an earlier argument based on a redundant field equation.
It is shown further that the stability condition applies to $k \ne 0$ Friedmann-Robertson-Walker spaces based on the non-redundant Friedmann equation derived from a simple effective Lagrangian.
We show how to derive this expression for the Friedmann equation of pure gravity theory.
This expression is also generalized to include scalar field interactions.
\end{abstract} \vskip .2in


\section{Introduction}
Inflationary theory provides an appealing resolution for the flatness, monopole, and horizon problems of our present universe described by the standard big bang cosmology \cite{inflation}. 
It is known that our universe is homogeneous and isotropic to high degree of precision \cite{data}.
Such a universe can be described by the well known Friedmann-Robertson-Walker (FRW) metric \cite{book}.
There are only three classes of FRW spaces characterized by their topological structure: one can either have a closed, open or flat universe according to the observations at large.

It is also known that gravitational physics should be different from the standard Einstein models near the Planck scale \cite{string,scale}. 
For example, quantum gravity or string corrections could lead to interesting cosmological consequences \cite{string}.
Moreover, some investigations have addressed the possibility of deriving inflation from higher order gravitational corrections \cite{Kanti99,s80,dm95}.

A general analysis of the stability condition for a variety of pure higher derivative gravity theories is very useful in many respects.  
It was shown that a stability condition should hold for any potential candidate of inflationary universe in the flat FRW space \cite{dm95,JB}. 
We will first briefly review the approach of reference \cite{dm95} based on a redundant field equation.
The proof will be shown to be incomplete.
We will also show how to complete the proof with the help of the Bianchi identity for some models where the redundant equation can be recast in a form similar to the Bianchi identity in a FRW background.

In addition, the derivation of the Einstein equations in the presence of higher derivative couplings is known to be very complicated.
The presence of a scalar field in induced gravity models and dilaton-gravity model makes the derivation even more difficult to derive.

We have developed a simpler derivation by imposing the FRW symmetry before varying the action, while keeping the proper time lapse function
\cite{kp91}.
We try to generalize the work in \cite{kp91} in order to obtain a general formula for the non-redundant Friedmann equation. 
It can be applied to provide an alternative and simplified method to prove the validity of the stability conditions in pure gravity theories.
In fact, this general formula for the Friedmann equation is very useful in many area of interests.

\section{Friedmann Equation and Stability of Pure Gravity Theories}

The generalized Friedmann-Robertson-Walker (GFRW) metric can be read off directly from the following equation:
\begin{equation}
ds^2 \equiv g^{\rm GFRW}_{\mu\nu} dx^\mu dx^\nu =  -b(t)^2dt^2 + {a^2}(t)
\Bigl( {dr^2 \over 1 - k
r^2} + r^2 d\Omega  \Bigr).
\label{eqn:frw} \label{GFRW}
\end{equation}
Here $ d \Omega $ is the solid angle
$d \Omega \, = \, d{\theta}^2 + {\sin}^2 \theta \, d{\chi}^2 ,$
and $k \, = \, 0, \pm 1$ stand for a flat, closed or open universe respectively.
Note also that the FRW metric can be obtained from the GFRW metric by setting the lapse function  $b(t)$ equal to one, i.e. $b=1$, in equation (\ref{eqn:frw}).

One can list all non-vanishing components of the curvature tensor as
\begin{eqnarray}
R^{ti}_{\,\,\,\,tj}&=&{1\over 2} [H\dot{B}+2B(\dot{H}+H^2)]\delta^i_j  ,
\label{Rti}\\
R^{ij}_{\,\,\,\,kl}&=& (H^2B+{k \over a^2} ) C^{ij}_{\,\,\,\,kl}   \label{Rkl}.
\end{eqnarray}
Here $C^{ij}_{\,\,\,\,kl} \equiv \epsilon^{ijm} \epsilon_{mkl}$ with
$\epsilon^{ijk}$ denoting the three space Levi-Civita tensor \cite{book}. 
Here $\dot{}$ denotes differentiation with respect to $t$ and
$H= \dot{a}/a$ is the Hubble constant.
We have written $B \equiv 1 / b^2$ for later convenience.

Given a pure gravity model one can cast the action of the system as 
$S=\int d^4x \sqrt{g} {\cal L}=N \int dt a^3L(H, \dot{H}, k/a^2)$ in the FRW spaces.
Here $N$ is a time independent integration constant. 
If we take $L$ as an effective Lagrangian, one can show that the variation with respect to $a$ gives 
\begin{eqnarray} \label{aeq}
3L- H {\delta L \over \delta H} + (H^2- \dot{H}) {\delta L \over \delta
\dot{H}} &=&
(2H + {d \over  dt})  [ -(4H + {d \over  dt})
{\delta L \over \delta \dot{H}}  +{\delta L \over \delta {H}} ]
+2k {\delta L \over \delta k} .
\end{eqnarray}
Note that $a^3L$ is normally referred to as the effective Lagrangian.
We will also call $L$ the effective Lagrangian unless confusion occurs.
The Above equation is the space-like $ij$ component of the Einstein equation
\begin{equation}
G_{\mu\nu} = t_{\mu\nu}          \label{einstein}
\end{equation}
with $t_{\mu\nu}$ denoting the generalized energy momentum tensor associated with the system.
It is known that this equation is in fact a redundant equation. 
Indeed, one can define 
$H_{\mu\nu} \equiv G_{\mu\nu} - t_{\mu\nu}$ and write the field equation as $H_{\mu\nu}=0$.

Hence one has
\begin{equation}
D_\mu H^{\mu\nu} =0 \label{DH}
\end{equation}
from the energy conservation ($D_\mu t^{\mu \nu}=0$) and the Bianchi identity ($D_\mu G^{\mu \nu}=0$).
Indeed, the extended Bianchi identity (\ref{DH}) can be shown to give
\begin{equation}
(\partial_t + 3 H) H_{tt} + 3 a^2H H_3=0, \label{h3}
\end{equation}
as soon as the FRW metric is substituted into equation (\ref{DH}).
Here $H_3 \equiv {1 \over 3} h^{ij} H_{ij}$ and $g_{ij} \equiv a^2 h_{ij}$.
It is now straightforward to show that $H_{ij} = H_3 h_{ij}$.
In fact, equation (\ref{h3}) indicates that: 
``$H_{tt}=0$ implies $H_3=0$" as long as  $a^2H \ne 0$. 
On the other hand, $H_3=0$ implies instead
$(\partial_t + 3H ) H_{tt}=0.$
This implies $a^3 H_{tt} = {\rm constant}.$
Hence the $_{ij}$ equation  can not imply the Friedmann equation
$H_{tt} =0$.
Hence any conclusion derived without the Friedmann equation is known to be incomplete.

We will briefly review the stability analysis obtained from the analysis based on the redundant equation (\ref{aeq}) \cite{dm95} here and show how to make up the loop-hole in this approach. 
Suppose that we are given a pure gravity theory, the stability of the background inflationary solution for the Hubble constant $H=H_0$ to the redundant field equation (\ref{aeq}) can be obtained by perturbing
$H=H_0+\delta H$. 
The leading order perturbation equation can be shown to be
\begin{equation}  \label{stadm}
3H_0 F + \dot{F}=0
\end{equation}
along with the zeroth order equation that vanish according to the field equation. 
This in fact takes a little arguments as shown in reference \cite{dm95}. 
One can show that the zeroth order perturbation equation from the perturbed Friedmann equation leads directly to the field equation for the background field.
For simplicity the parameter $k$ is set as $k=0$ in reference \cite{dm95}.
Here $F$ is  defined as
\begin{equation}
F\equiv L_{22}(0)\delta  \ddot{H}+3H_{0}L_{22}(0)\delta
\dot{H}+(6L_{2}(0)+3H_{0} L _ { 2 1 } (0)-L_{11})\delta H  \label{stable}
\end{equation}
In addition, the coefficient of expansions are defined by
\begin{eqnarray}
L(H,\dot{H})
&=& L(H_{0},0)+\frac{\delta L}{\delta
H}(H_{0},0)\delta H+\frac{\delta
L}{\delta \dot{H}}(H_{0},0) \delta \dot{H}
\equiv  L(0)+L_{1}(0)\delta H+L_{2}(0)\delta \dot{H}
\\
\frac{\delta L}{\delta H}(H,\dot{H})
&=&\frac{\delta L}{\delta
H}(H_{0},0)+\frac{\delta^{2}L}{(\delta
H) ^{2}}(H_{0},0)\delta  H+\frac{\delta^{2}L}{\delta
H\delta\dot{H}}(H_{0},0)\delta \dot{H}
\equiv L_{1}(0)+L_{11}(0) \delta H+L_{12}(0)\delta  \dot{H}
\\
\frac{\delta L}{\delta \dot{H}}(H,\dot{H})
&=&\frac{\delta L}{\delta \dot{H}}(H_{0},0)
+\frac{\delta^{2}L}{\delta H \delta\dot{H} }(H_{0},0) \delta H
+\frac{\delta^{2}L}{(\delta \dot{H})^{2}}(H_{0},0)\delta \dot{H}
\equiv  L_{2}(0)+L_{21}(0)\delta H +L_{22}(0)\delta  \dot{H}
\end{eqnarray}
If we focus on the solution $F=0$ \cite{dm95}, one has 
\begin{equation}
\delta H = A_+ e^{B_+t} +A_- e^{B_-t}.
\end{equation}
Here $A_{\pm}$ denotes arbitrary constants and
\begin{equation}
B_{\pm} = - {3 \over2} H_0 \pm {\sqrt{\Delta} \over 2 L_{22}} \label{bpm}
\end{equation}
denotes the characteristic roots to the characteristic equation
\begin{equation}
L_{22}(0) x^2+3H_{0}L_{22}(0)x + 6L_{2}(0)+3H_{0}
L _ { 2 1 } (0)-L_{11} =0 \label{char}
\end{equation}
of the ODE (\ref{stable}).
Here $\Delta \equiv 9H_0^2L_{22}^2 -4L_{22}(6L_2 +3H_0L_{21} -L_{11})$
denotes the discriminant of the characteristic equation of (\ref{stable}).

One can integrate $\delta H$ to obtain
\begin{equation}  \label{condition}
a(t)= a_0 \exp \left( {H_0t + {A_+ \over B_+} e^{B_+t} +
{A_- \over B_-} e^{B_-t}} \right) .
\end{equation}
%if the system is stable against the perturbation $\delta H$.
Therefore, one finds that stability of the de Sitter type inflationary
solution will require both characteristic roots $B_{\pm}$
to be negative. If one of the roots is positive and the other one is negative,
then
there may exist a limited period of inflation.
This sort of inflation will come to an end
in a time duration of the order of ${1 / B_p}$ with $B_p$ denoting the
positive root.  Choosing a sufficiently small value of $1/B_p$ allows
inflation to exit naturally \cite{dm95}.
Therefore the
sign of the roots to the characteristic equation (\ref{char})
can be checked to see if the system supports a stable inflationary de Sitter
solution. If the discriminant is negative, the solution of $B_{\pm}$ will
contain an oscillating phase. Hence the system is stable again.
Since this argument is based on the redundant field equation, this
stability analysis is not complete. In other words,
the redundant $G_{ij}$ equation will normally take the form of $\partial_t(a^3
G_{tt})=0$. Hence analysis based on the $G_{ij}$ equation will be quite indirect
and incomplete.
%In addition, our analysis also applies to $k \ne 0$ cases if we are interested
%in the stability of the inflationary solution such that $|\dot{a}| \gg 1$.

There are two problems with this stability condition.
First of all, this condition is obtained from the redundant equation.
One does not know the validity of the field equation, not to mention the
stability condition derived from it.
Secondly, there are homogeneous terms in equation (\ref{stadm}) in addition
to $F=0$, i.e. $F=k_1 \exp [-3H_0t]$ with an arbitrary constant $k_1$. The first
problem is not easy to answer for the moment.
The second problem can be resolved immediately. One notes that the complete
solution to the redundant equation (\ref{stadm}) is in fact
$\delta H = A_+ e^{B_+t} +A_- e^{B_-t} +k_1/[
(6L_{2}(0)+3H_{0} L_{ 2 1 } (0)-L_{11})a_0^3]$. Here $a_0(t)= \exp{H_0t}$.
This obviously will not
affect the stability analysis as long as we are interested in the inflationary
universe where the particular solution is negligible in the above equation
unless the denominator of the $k_1$-term happens to vanish.
In fact, we are going to show that $F=0$ is not only a lucky guess, it can be derived from perturbing the Friedmann equation.
But one can not be sure about this unless a closed form expression for the
Friedmann equation is available so that a model independent
analysis is applicable.

Nevertheless, one can still resolve this problem by looking into
the details of the Bianchi identity.
As to the first problem with this condition, one notes that
in most cases, the redundant equation can be rearranged as 
\begin{equation} \label{d3h}
\partial_t (a^3 H_{tt})=0
\end{equation}
using the Bianchi identity. The solution to above equation is $H_{tt}=
{\rm constant}
\times a^{-3}$. Hence one can show that the Friedmann equation has to be of
the form
\begin{equation} \label{d3f}
H_{tt}=\tilde{F}+k_1 a^{-3}=0
\end{equation}
if the redundant equation can be written as the combination $\partial_t (a^3 \tilde{F})=0$ with
$\tilde{F}=0$ the corresponding equation leading to the first order equation $F=0$ shown in Ref. \cite{dm95}.
To be more specifically, $\delta \tilde{F} =F +$ to the leading order in $\delta H$ and its derivatives.
Here $H=H_0+\delta H$.
This follows from the fact that $\partial_t [a^3 (H_{tt}-\tilde{F})]=0$ implies that the difference $H_{tt}-\tilde{F}$ has to be proportional to
$a^{-3}$ with some arbitrary constant $k_1$.
Therefore, one can effectively work with the $F=0$ solution
if we are working on an inflationary background De Sitter solution.
This is because $H_{tt} \simeq \tilde{F}$ in the De Sitter background.
Therefore, any analysis based on the ansatz $F=0$ can only be justified in the
De Sitter background. In particular, stability conditions derived from
$F=0$ adopted in Ref. \cite{dm95} can not be justified from above analysis
in anti De Sitter space. This is because the undetermined part $k_1
a^{-3}$ will affect the result significantly.

While we suspect that $F=0$ should probably be the first order Friedmann
equation we are
looking for, we are not sure if the redundant equation can always be cast
into the familiar form shown above.
Moreover, the true Friedmann equation can look like
$\tilde{F}+k_1/a^3=0$ even if we can write the redundant equation  in above familiar
form.
Fortunately, one can in fact derive a closed form for the Friedmann
equation similar to equation (\ref{aeq}).

%\section{Friedmann Equation}
The Friedmann equation can be recast as
\begin{equation}
L + (H {d \over dt}+ 3H^2-\dot{H}) {\delta L \over \delta \dot{H}} -
                           {\delta L \over \delta H} H =0  \label{key}
\end{equation}
after some algebra. This is done by a variation of $L^{\rm GFRW}$ with respect
to $b$
(or equivalently with respect to $g_{tt}$) and setting $b=1$ afterwards. Here
$L^{\rm GFRW}\equiv \int d^3x {\cal L} (g_{\mu \nu} = g^{\rm GRFW}_{\mu \nu})$.
One
notes that the crucial point in the derivation is due to the observation that
any variation of $L$ with respect to $H\dot{B}$ has to be equivalent to the
variation of $L$ with respect to $2B\dot{H}$. This is because the term
$H\dot{B}$ always shows up with $2B\dot{H}$ as indicated in the explicit
formulae listed in equations (\ref{Rti}-\ref{Rkl}).
Note that equation (\ref{key}) is known as the minimum Hamiltonian constraint
${\cal H} \equiv \pi \dot{a} -L(a, \dot{a})=0$ in the case where ${\delta L
/ \delta \dot{H}}=0$. For example, one can write
$L=-R=6[ k/a^2\,\, -H^2 ]$ after proper integration by parts.
Hence the Hamiltonian constraint is identical to the Friedmann equation
(\ref{key}) in this model.

Note in particular that even the term containing $k$ does not get involved
explicitly in the Friedmann equation, equation (\ref{key}) remains valid for arbitrary $k$.
Our derivation leading to equation (\ref{key}) is based
on a pure gravitational action.
The derivation of the
Friedmann equation in the presence of other sources of interactions is
straightforward.
In addition, the Friedmann equation in $D$-dimensional FRW
space \cite{dm2} can also be derived following similar arguments.

One can then apply the same perturbation,
$H=H_0+ \delta H$, to the Friedmann equation.
The zeroth order perturbation equation gives exactly
the field equation for the background field $H=H_0$ while
the leading order in $\delta H$ gives $F=0$ identically. 
Indeed, perturbing equation (\ref{key}) gives
\begin{equation}
L_{22}(0)\delta  \ddot{H}+3H_{0}L_{22}(0)\delta
\dot{H}+(6L_{2}(0)+3H_{0} L _ { 2 1 } (0)-L_{11})\delta H =0 \label{stable1}
\end{equation}
to the leading order in $\delta H$ and its derivatives.
Note that $a$-dependent terms always appear in a combination as $H^2 + k/a^2$ in $R^{ij}_{\:\: kl}$ as given by equation (\ref{Rkl}).
Hence one can ignore the $\delta a$-dependent terms during the inflationary phase when $H \gg 1/a^2$.
This follows from the fact that $\delta a  \sim a \delta H \Delta t$ and hence $|\delta (1/a^2) / \delta H^2| \sim 1/Ha^2 \ll 1$ during the inflationary phase.
Here $\Delta t$ is the time duration for the inflationary phase.
Therefore, one
is able to show that the stability conditions in the inflationary phase
are indeed given by the result obtained in equation (\ref{condition}).
Hence, one would never need to worry about any complication that
can possibly
weaken the validity of the stability condition obtained in reference
\cite{dm95}. This stability condition hence serves as a screening device for any
possible candidates for a realistic cosmological universe without any ambiguity.
%In addition, the stability condition obtained directly from the Friedmann equation remains valid for $k\ne 0$ FRW spaces. 
In addition, the stability condition obtained earlier works also for curved FRW spaces in the inflationary phase
where $H \gg {1 / a^2}$.

In short, our result states clearly without ambiguity that physically
acceptable inflationary
models need to meet the stability conditions shown earlier in this section. The perturbative stability indicates that a solution with a stable mode and an unstable mode can possibly exit the inflationary phase in due time. 
Our result based on
the non-redundant Friedmann equation is complete
and remains valid for all FRW models.
Most of all, working directly on the Friedmann equation (\ref{key}) we just
derived can save us a lot of trouble in the complete analysis.

\section{Friedmann Equation for Scalar-Gravity Theory}
The derivation of the Friedmann equation in the presence of a scalar field
is in fact rather straightforward since complications only arise from
complicated curvature interactions.
Indeed, the inclusion of scalar interactions introduces a kinetic term
$T_\phi = -{1 \over 2} \partial_\mu \phi \partial_\nu \phi g^{\mu \nu} .$
%\label{Tphi} \end{equation}
It will take the form
$T_\phi = {1 \over 2} \dot{\phi}^2 B(t)$
if $\phi({\bf x}, t) =\phi(t)$.
Hence the complete effective Lagrangian in the GFRW spaces will take the
following form
\begin{equation}
ba^3L(g^{GFRW},\phi)= ba^3L_0 +{1 \over 2} a^3 \sqrt{B} \dot{\phi}^2 ,
\end{equation}
with $L_0$ denoting the graviton Lagrangian plus everything else except
the kinetic term of the scalar field $T_\phi$. Or equivalently,
$L_0=L-T_\phi$ with $L$ denoting the complete effective Lagrangian of the
theory.
Hence, one can show that the
Friedmann equation becomes
\begin{equation}
L_0 - T_\phi + (H {d \over dt}+ 3H^2-\dot{H}) {\delta L_0 \over \delta \dot{H}}
-  {\delta L_0 \over \delta H} H =0  \label{keyp}         .
\end{equation}
Note that the minus sign in front of $T_\phi$ is due to the $a^3b^{-1}L$
combination from the $\sqrt{g}$ and the $g^{tt}$ component.
In addition, the variational equation for the $\phi$ field can be
directly obtained
from the variation of the effective Lagrangian $L$ with respect to $\phi$.

The method for deriving the Friedmann equation described here can be
extended to theories with any form of simple gravitational interactions in a
straightforward way.
For example, one can study the following action with Gauss-Bonnet coupling
\cite{Kanti99}
\begin{equation}
L=-\frac{1}{2} R - \frac{1}{2}\partial_\mu \phi \partial^\mu \phi
+f(\phi)R_{GB}^{2}
\end{equation}
with $R_{GB}^{2}=
R_{\mu\nu\alpha\beta}R^{\mu\nu\alpha\beta}-4R_{\mu\nu}R^{\mu\nu}+R^2$
denoting the Gauss-Bonnet term. The effective Lagrangian
is then
\begin{equation}
L= 3 ( \dot{H} + 2 H^2 +{k \over a^2} ) + {1 \over 2} \dot{\phi}^2
+24 (\dot{H} + H^2)(  H^2 +{k \over a^2} ) f(\phi)
\end{equation}
once the FRW metric is applied.
The Friedmann equation (\ref{keyp}) becomes
\begin{equation}
3(H^{2}+\frac{k}{a^{2}})(1+8H\dot{f})=\frac{\dot{\phi}^{2}}{2}
\end{equation}
Furthermore, the variational equation of $\phi$ is also straightforward.
The result is
\begin{equation}
\ddot{\phi}+3\dot{\phi}H-
24\frac{df}{d\phi}(\dot{H}+H^{2})(H^2+\frac{k}{a^{2}})=0 .
\end{equation}
This agrees with the result in \cite{Kanti99} while the derivation is much
more straightforward. In fact this simple formula for the Friedmann equation can
also be generalized to any scalar-gravity theory. It helps to reduce the labor
in deriving gravitational field equations. It is especially helpful when
complicated interactions are present and higher derivative terms become
important.

\vspace{0.5in}
\section{Conclusions and Acknowledgments}
 
A general analysis of the stability condition for a variety of pure higher
derivative gravity theories is very useful in many respects. 
It is known that a stability condition should hold for any potential candidate of inflationary universe in the flat FRW space \cite{dm95}. 
We first briefly review the approach of reference \cite{dm95} based on a redundant field equation.
The proof is shown to be incomplete in this paper.
We also showed in this paper how to complete the proof with the help of the Bianchi identity for some models where the redundant equation can be recast in a form similar to the Bianchi identity in a FRW background.
 
In addition, the derivation of the Einstein equations in the presence of
higher derivative couplings is known to be very complicated.
For example, the presence of a scalar field in induced gravity models and dilaton-gravity model makes the derivation even more difficult to derive. 
 
We have developed, in this paper, a simpler derivation by imposing the FRW symmetry before varying the action, while keeping the lapse function.
We also tried to generalize the work in \cite{kp91} in order to obtain a general formula for the non-redundant Friedmann equation in this paper.
This result was applied to provide an alternative and simplified method to prove the validity of the stability conditions in pure gravity theories.
This general formula for the Friedmann equation is also very useful in many area of interests. 
 
This work was supported in part under the contract number NSC88-2112-M009-001 and NSERC grant 72013704.

%\section{Acknowledgments}
%{\bf \large Acknowledgments}
%This work was supported in part under the contract number
%NSC88-2112-M009-001 and NSERC grant 72013704.

\end{document}